\documentclass[letterpaper, 10 pt, conference]{ieeeconf}  

\IEEEoverridecommandlockouts                              

\usepackage[T1]{fontenc}
\usepackage{graphicx} 
\usepackage{amsmath} 
\usepackage{amssymb}  
\usepackage{mathtools}
\usepackage[hyphens]{url}

\usepackage[ruled,vlined,linesnumbered]{algorithm2e}
\usepackage{algorithmic,algorithm2e,float}

\SetKw{Continue}{continue}

\usepackage{amsthm}
\theoremstyle{definition}
\newtheorem{problem}{Problem}

\newtheorem{definition}{Definition}

\newtheorem*{rem}{Remark}

\newcommand\oprocendsymbol{\hbox{$\bullet$}}
\newcommand\oprocend{\relax\ifmmode\else\unskip\hfill\fi\oprocendsymbol}

\newcommand{\sss}{Sparse-SS}
\newcommand{\ipp}{IPP}

\title{\LARGE \bf Informative Path Planning in Random Fields via \\Mixed Integer Programming}

\author{Shamak Dutta, Nils Wilde, and Stephen L. Smith
\thanks{This research is supported in part by the Natural Sciences and Engineering
Research Council of Canada (NSERC) and by Nutrien Ltd.}
\thanks{S.\ Dutta and S.\ L.\ Smith are with the Department of Electrical and Computer Engineering,
        University of Waterloo, Canada 
        \{{\tt\small stephen.smith, shamak.dutta\}@uwaterloo.ca}.  N.\ Wilde is with the Cognitive Robotics Department, Delft University of Technology, Netherlands ({\tt\small N.Wilde@tudelft.nl}).}%
}
\begin{document}
\maketitle
\thispagestyle{empty}
\pagestyle{empty}

\begin{abstract}
We present a new mixed integer formulation for the discrete informative path planning problem in random fields. The objective is to compute a budget constrained path while collecting measurements whose linear estimate results in minimum error over a finite set of prediction locations. The problem is known to be NP-hard. However, we strive to compute optimal solutions by leveraging advances in mixed integer optimization. Our approach is based on expanding the search space so we optimize not only over the collected measurement subset, but also over the class of all linear estimators. This allows us to formulate a mixed integer quadratic program that is convex in the continuous variables. The formulations are general and are not restricted to any covariance structure of the field. In simulations, we demonstrate the effectiveness of our approach over previous branch and bound algorithms.
\end{abstract}


\section{Introduction} \label{section:introduction}
Consider the following problem. Physical processes such as temperature or soil nutrient variability exhibit variation over large domains. A robot is tasked with collecting measurements in this environment to build an accurate map of the process. With unlimited resources, a dense sampling strategy yields good results. However, robots have constraints such as battery life, fuel capacity, or maximum path length. The challenge is to plan budget constrained paths while collecting observations to maximize the information (or equivalently minimize the estimation error) in the environment. This is known as the \emph{informative path planning problem} (IPP).

\begin{figure}
    \centering
    \includegraphics[width=0.99\linewidth]{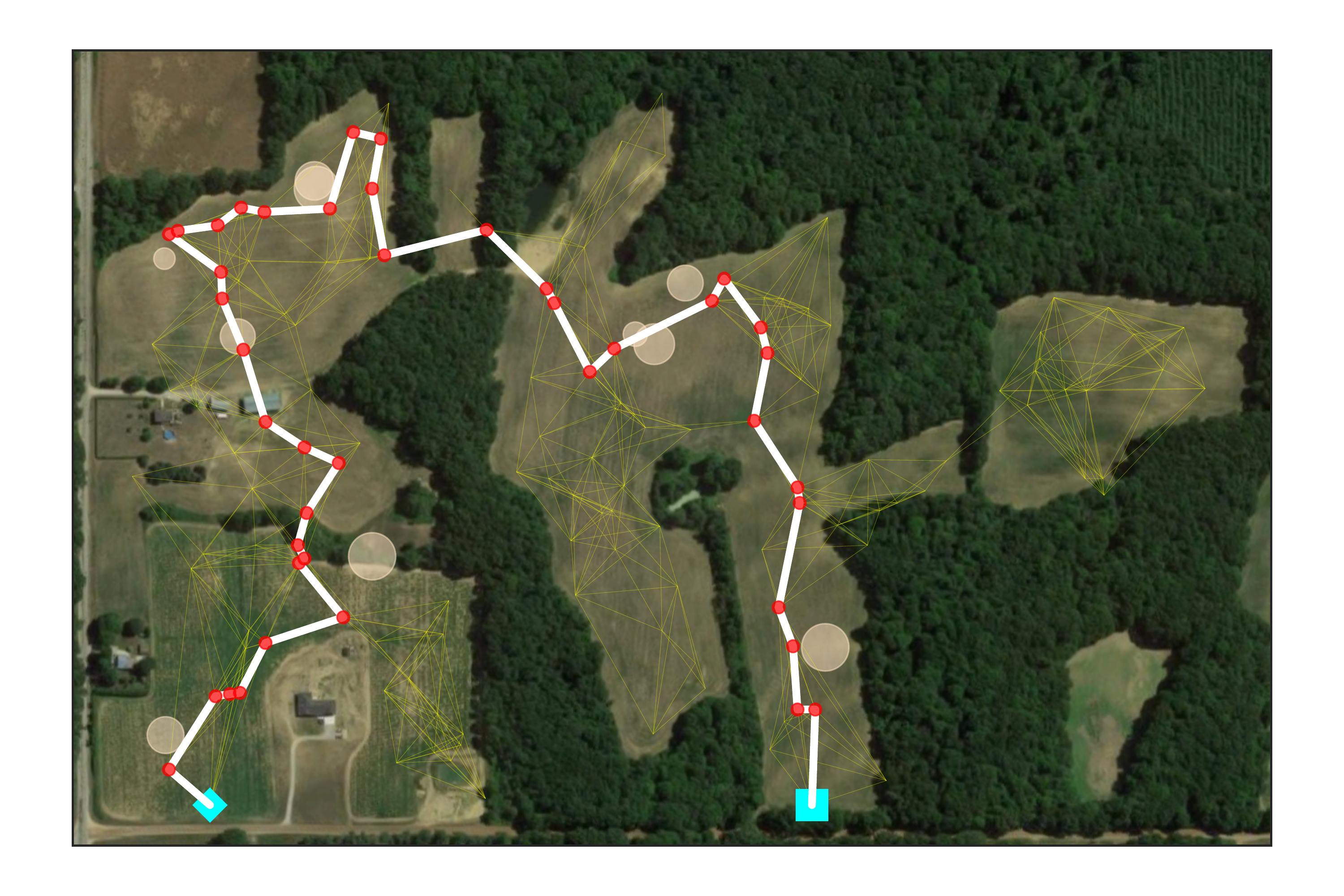}
    \caption{An example of an informative path planning problem. The robot must plan a budgeted path from start to end (light blue boxes) and maximize the prediction accuracy at the beige circles using observations on its path (red points). The area of the beige circles represents its importance.}
    \label{figure:intro_figure}
\end{figure}

We address the informative path planning problem in environments modeled as random fields. This framework is powerful because estimates at \emph{any} unobserved location can be computed from a set of measurements. In addition, the expected estimation error can be quantified \emph{a priori}. The setup is as follows: we are given a set of prediction and observation variables associated with locations in an environment. Given a budget $B$, the objective is to compute a path of length at most $B$ that minimizes the estimation error over the prediction variables (see Figure~\ref{figure:intro_figure}). The dual formulation is to plan minimum length paths while ensuring the estimation error is within a given tolerance.  

In this paper, we assume that we are given $N$ prediction locations and $M$ observation locations associated with a $d$-dimensional random field whose covariance structure is known. The estimation quality of a measurement set is evaluated using the mean squared error resulting from the linear least-squares estimator. Though informative path planning is known to be NP-hard in this setting, our objective is to develop an approach to computing optimal solutions.

\emph{Contributions:} We give a new mixed integer formulation for the informative path planning problem with the objective of minimizing the estimation error in a random field. Our formulation has several properties. First, the objective is a convex quadratic function in the continuous variables making it amenable to modern optimization solvers. This allows us to tackle larger problem instances not previously considered by optimal solvers in the literature. Second, the formulation is general and is not restricted to any covariance structure of the random field. Third, MIP solvers provide lower bounds on the estimation error if terminated early, which can be used to provide suboptimality certificates for approximate solutions. The key idea of our approach is to expand the search space so we optimize not only over the measurement subset, but also over the class of all linear estimators. While there is no guarantee on runtime, we demonstrate that our approach provides benefits both in terms of solution quality and runtime over previous branch and bound algorithms for informative path planning.

\emph{Related Work:} Informative path planning is closely related to research in sensor placement as the objectives are often similar. In sensor placement, the goal is to choose the best locations to deploy a set of sensors such that the information or sensing quality is maximized.  Greedy strategies that maximize mutual information \cite{krause2008near} have yielded approximation guarantees due to submodularity but do not provide guarantees on the estimation error. Other objectives such as robustness and resiliency \cite{tzoumas2017resilient} and estimation error in dynamical systems \cite{dhingra2014admm,kohara2020sensor,tzoumas2016sensor} have been studied. Approximation algorithms that provide guarantees on the estimation error have been studied in the case of continuous environments \cite{suryan2020learning}. A related problem appears in subset selection for regression where the challenge is to select $k$ random variables that best predict another variable. Greedy algorithms are popular for minimizing the estimation error \cite{das2008algorithms} due its computational efficiency and strong empirical results. The work in \cite{das2008algorithms} identifies special problem instances where approximation guarantees are provided. The MIP formulation in this paper is inspired by the proof of hardness of subset selection for linear regression in \cite{das2011subset}, which performs a reduction from the NP-hard sparse approximation problem~\cite{natarajan1995sparse}. By leveraging recent advances in mixed integer optimization for sparse approximation \cite{bertsimas2016best}, we compute optimal solutions to instances informative path planning not considered previously.

In informative path planning, an important step is deciding the maximally informative locations to observe. However, there is the added constraint of path cost which prevents sampling at all informative locations. A recursive greedy approach \cite{chekuri2005recursive} used in environmental monitoring \cite{binney2010informative,binney2013optimizing} provides guarantees when the objective is submodular but runs in quasi-polynomial time, limiting its practical applicability. Recent work in adaptive sampling for environmental monitoring considers the estimation error (or equivalently variance reduction) as an information gain function \cite{chen2019multi}. Our objective does not depend on the outcome of measurements and can be computed \emph{a priori}. We benchmark our approach against branch and bound algorithms \cite{binney2012branch} which are computationally expensive on moderately sized graphs. Mixed integer programs have been proposed to solve the correlated orienteering problem~\cite{yu2016correlated} where the reward is a designed quadratic utility function capturing the spatial correlation. In contrast, our work considers the estimation error in random fields. Orienteering~\cite{chao1996fast} is concerned with finding a budget constrained tour in a graph that maximizes the reward collected at vertices. A formulation where a certain reward must be collected in minimum time is considered in \cite{sadeghi2019minimum}. In contrast, our work considers a general cost function on a subset of vertices namely, the estimation error.

\emph{Organization:} In Section \ref{section:preliminaries}, we review orienteering and linear least-squares estimation, the latter of which is central to the main idea of the paper. In Section \ref{section:problem_formulation}, we formalize the problem of subset selection and informative path planning in random fields. In Section \ref{section:mixed_integer_formulation}, we describe our solution approach and provide the MIPs. Finally, in Section \ref{section:results}, we demonstrate the effectiveness of our algorithm by comparing against previous branch and bound techniques.
\section{Preliminaries} \label{section:preliminaries}
In this section we review linear least-squares estimation \cite{kumar2015stochastic} and orienteering in graphs \cite{gunawan2016orienteering}.

\subsection{Linear Least-Squares Estimation}\label{subsection:least_squares_estimation}
Let $X_1, \ldots, X_n, Y$ be square integrable zero mean random variables. Define $\boldsymbol{b} := (\text{Cov} (X_1, Y), \ldots, \text{Cov} (X_n, Y))' \in \mathbb{R}^n$, $\boldsymbol{X} := (X_1, \ldots, X_n)'$ and let $C \in \mathbb{R}^{n \times n}$ where $C_{ij} = \text{Cov}(X_i, X_j)$. The linear least-squares estimator of $Y$ given $X_1, \ldots, X_n$ is given by the following definition.  Note, we include a detailed expression as it will be used later in our mixed integer formulations.
\begin{definition}[Linear Least-Squares Estimator] \label{definition:linear_least_squares_estimator}
Given $X_1, \ldots, X_n$, the optimal linear estimator, in the least-squares sense, of $Y$ is
\begin{equation}
\hat{Y} := \boldsymbol{\alpha}_*' \boldsymbol{X},
\end{equation}
where the optimal coefficient vector $\boldsymbol{\alpha}_* \in \mathbb{R}^n$ is the solution to the following convex quadratic function,
\begin{equation}
    \begin{split}
        \boldsymbol{\alpha}_* &= \underset{\boldsymbol{\alpha}}{\text{arg min }} \mathbb{E} ( (Y - \boldsymbol{\alpha}' \boldsymbol{X})^2 )\\
        &= \underset{\boldsymbol{\alpha}}{\text{arg min }} \langle Y - \boldsymbol{\alpha}' \boldsymbol{X},   Y - \boldsymbol{\alpha}' \boldsymbol{X} \rangle\\
        &= \underset{\boldsymbol{\alpha}}{\text{arg min }} \boldsymbol{\alpha}' C \boldsymbol{\alpha} - 2 \boldsymbol{b}' \alpha + \text{Cov}(Y, Y)\\
        &= C^{-1} \boldsymbol{b}.
    \end{split}
\end{equation}
\end{definition}

The estimation error is given by the following definition.
\begin{definition}[Mean Squared Estimation Error] \label{definition:mean_squared_error}
Given $X_1, \ldots, X_n$, the linear least squares estimator of $Y$ results in mean squared estimation error given by
\begin{equation}
\mathbb{E}((Y - \hat{Y})^2) := \text{Cov}(Y,Y) - \boldsymbol{b}^T C^{-1} \boldsymbol{b}.
\end{equation}
\end{definition}

\subsection{Generalized Orienteering in Graphs} \label{subsection:orienteering}
The input to the general orienteering problem is a directed graph $G=(V,A)$, two nodes $s,t \in V$, a budget $B > 0$, and a reward function $R: 2^V \rightarrow \mathbb{R}$.  An $s$-$t$ path $P$ in $G$ is a sequence of distinct vertices $\langle s, v_1, \ldots, v_n, t \rangle$. With slight abuse of notation, we let $P$ refer to the path as well as the set of vertices visited on the path.  The goal is then to find a $s$-$t$ path $P$ of length at most $B$ such that the reward $R(P)$ is maximized. Maximizing the reward function can be replaced with minimizing a cost function $C: 2^V \rightarrow \mathbb{R}_{\geq 0}$ where the cost is a monotonically decreasing set function.

\section{Problem Formulation} \label{section:problem_formulation}
Consider an environment $D \subset \mathbb{R}^d$, $d \in \mathbb{Z}_{>0}$, and a random field $\{Z(x): x \in D\}$, where for each $x \in D$, $Z(x)$ is a zero-mean random variable with finite variance. Define $\phi: \mathbb{R}^d \times \mathbb{R}^d \rightarrow \mathbb{R}_{\geq 0}$ to be the covariance function associated with the random field i.e.,\ for any $x, y \in D$,
\begin{equation}
    \text{Cov}\left( Z(x), Z(y) \right) = \mathbb{E} \left(Z(x) Z(y)\right)= \phi(x, y).
\end{equation}

We assume the random field can be observed, subject to measurement noise. Specifically, for some point $x \in D$, the measurement equation is
\begin{equation}
    Y(x) := Z(x) + \epsilon(x),
\end{equation}
where $\epsilon(x)$ is a zero-mean random variable with variance $\sigma^2 > 0$. The measurement noise is assumed to be uncorrelated with the process $Z(x)$ and uncorrelated across environment locations i.e.,\ for any $x,y \in D$, $\text{Cov}\left(\epsilon(x), \epsilon(y)\right) = 0$.

Given a measurement set $S = \{x_1, \ldots, x_n\} \subset D$, the linear least-squares estimate of $Z(x)$ is a linear combination of the observations in $S$,
\begin{equation}
    \hat{Z}(x) := \sum_{i=1}^n \alpha_i Y(x_i) = \boldsymbol{\alpha}_{*}' \boldsymbol{Y}_S,
\end{equation}
where the optimal coefficients are given by Definition \ref{definition:linear_least_squares_estimator} and the following notation is used:
\begin{equation}
    \begin{split}
        \boldsymbol{Y}_S &:= [Y(x_1), \ldots, Y(x_n)]' \in \mathbb{R}^n\\
        \boldsymbol{b}_{x,S} &:= [\phi(x,x_1), \ldots, \phi(x,x_n)] \in \mathbb{R}^n\\
        C_S &:= \mathbb{E} \left[ \boldsymbol{Z}_S \boldsymbol{Z}_S^T\right] + \sigma^2 I_n \in \mathbb{R}^{n \times n}\\
        &= \begin{bmatrix}
        \phi(x_1, x_1) & \dots & \phi(x_1, x_n)\\
        \vdots & \ddots & \vdots\\
        \phi(x_n, x_1) & \dots & \phi(x_n, x_n)
        \end{bmatrix} + \sigma^2 I_n\\
        \boldsymbol{\alpha_*} &:= [\alpha_1, \ldots, \alpha_n]' =  C_{S}^{-1} \boldsymbol{b}_{x,S} \in \mathbb{R}^n.
    \end{split}
\end{equation}
The resulting estimation error is given by Definition \ref{definition:mean_squared_error},
\begin{equation} \label{equation:estimation_error}
    \begin{split}
         f_x(S) &:= \mathbb{E}\left((Z(x) - \boldsymbol{\alpha}_*' \boldsymbol{Y}_S)^2\right)\\
         &= \text{Cov}\left(Z(x), Z(x)\right) - \boldsymbol{b}_{x,S}' C_S^{-1} \boldsymbol{b}_{x,S}\\
         &= \phi(x, x) - \boldsymbol{b}_{x,S}' C_S^{-1} \boldsymbol{b}_{x,S}.
    \end{split}
\end{equation}
When the random variables in the field are jointly normally distributed, the mean squared estimation error $f_x(S)$, for some $x \in D$, is known as the \textit{posterior variance} in Gaussian Process regression or the \textit{kriging variance} in geostatistics.

Now, we define the problem inputs. Consider a set of $M$ observation points $\Theta = \{y_1, \ldots, y_M\} \subset D$, a set of $N$ prediction points $\Omega = \{x_1, \ldots, x_N\} \subset D$, and a weight function $w: \Omega \rightarrow \mathbb{R}_{\geq 0}$ assigning an importance to each element in the prediction set $\Omega$. The observation locations are represented as a directed graph $G=(\Theta,A,d)$ where the vertex set is the set of observation locations $\Theta$. For each arc $a = (u, v) \in A$, the arc cost is given by $d(u, v)$. Note that the prediction set may be disjoint from the observation set.

The goal is to develop a general approach to solve problems closely linked to each other by the estimation error. The first relates to subset selection (SS) and the second relates to informative path planning (IPP).

\begin{problem}[\textsc{\sss}]
Given $\Theta$, $\Omega$, $w(\cdot)$, and $k > 0$, find a measurement set $S \subset \Theta$ of size $k$ such that the total weighted estimation error $\sum_{x \in \Omega} w(x) f_x(S)$ is minimized.
\end{problem}


\begin{problem}[\textsc{\ipp}]
Given $G=(\Theta, E, d)$, $\Omega$, $w(\cdot)$, $B > 0$, and nodes $s, t \in \Theta$, compute an $s$-$t$ path $P$ of length at most $B$ such that the total weighted estimation error $\sum_{x \in \Omega} w(x) f_x(P)$ is minimized.
\end{problem}

\emph{Error Constrained Variants of Problems:} The error constrained version of the subset selection problem involves finding a minimum cardinality measurement set whose maximum weighted estimation error over $\Omega$ is within a prescribed error tolerance. Similarly, the path planning version is to minimize path length while ensuring the estimation error constraint. Our approach requires slight modifications to accommodate these problems so we focus on the formulations for \textsc{\sss} and \textsc{\ipp}.


\section{Solution Approach} \label{section:mixed_integer_formulation}

The goal of this paper to optimally solve practical instances of subset selection and informative path planning in random fields. We present our approach to the reformulation of the estimation error. This allows us to pose MIPs for the problems described in Section \ref{section:problem_formulation}.

\subsection{Starting with \textsc{\sss}} \label{subsection:sparse_subset_selection}
In this subsection, we explain the main idea of this paper by first tackling \textsc{\sss}. The problem formulation of \textsc{\sss} is
\begin{equation} \label{equation:opt_sss_initial}
    \begin{aligned}
        & \underset{S \subset \Theta}{\text{minimize}}
        & & f(S)\\
        & \text{subject to}
        & & |S| = k,
    \end{aligned}
\end{equation}
where 
\begin{equation} \label{equation:total_estimation_error}
    \begin{split}
        \Omega &= \{x_1, \ldots, x_N\}\\
        f(S) &:= \sum_{i=1}^N w(x_i) f_{x_i}(S)\\
        f_{x_i}(S) &= \phi(x_i, x_i) - \boldsymbol{b}_{x_i,S}' C_S^{-1} \boldsymbol{b}_{x_i,S}.
    \end{split}
\end{equation}
 
One idea is to model \eqref{equation:opt_sss_initial} as an integer linear program (ILP). 
This seems like a reasonable approach since there are a finite number of measurement locations to choose from. The setup would involve binary decision variables for each measurement location that encode whether the measurement is selected. In addition, the cardinality constraint in \eqref{equation:opt_sss_initial} can be easily formulated. However, the trouble lies in modeling the objective $f(S)$ as a linear function of the binary decision variables. As seen in Equation \eqref{equation:total_estimation_error}, the total estimation error $f(S)$ is a function of $f_x(S)$ which is a non-linear and in general, a non-convex function of the measurement set. The linearization of $f(S)$ would enable the formulation of an ILP but the formulation is not immediately obvious.

However, there is a way forward. Recall from Definitions \ref{definition:linear_least_squares_estimator} and \ref{definition:mean_squared_error} that the least-squares estimation error results from from the \textit{optimal} linear estimator i.e.,
\begin{equation}
    f_{x_i}(S) = \underset{\boldsymbol{\alpha}_i \in \mathbb{R}^{|S|}}{\min }\ \boldsymbol{\alpha}_i' C_S \boldsymbol{\alpha}_i - 2 \boldsymbol{b}_{x,S}' \boldsymbol{\alpha}_i + \phi(x_i,x_i).
\end{equation}
Then, the total estimation error is
\begin{equation}
    \begin{split}
        f(S) &= \sum_{i=1}^N\ w(x_i) \left( \underset{\boldsymbol{\alpha}_i}{\min }\ \boldsymbol{\alpha}_i' C_S \boldsymbol{\alpha}_i - 2 \boldsymbol{b}_{x,S}' \boldsymbol{\alpha}_i + \phi(x_i,x_i) \right)\\
        &= \underset{\substack{\boldsymbol{\alpha}_1, \\
        \ldots, \\
        \boldsymbol{\alpha}_N}}{\min }\ \sum_{i=1}^N\ w(x_i) \left( \boldsymbol{\alpha}_i' C_S \boldsymbol{\alpha}_i - 2 \boldsymbol{b}_{x,S}' \boldsymbol{\alpha}_i + \phi(x_i,x_i) \right).
    \end{split}
\end{equation}
This is the key to our formulation. With this observation, we can augment the original problem in \eqref{equation:opt_sss_initial} to optimize over both the measurement set $S$ and the coefficients $\boldsymbol{\alpha}_i$.  Note that each prediction variable $x_i \in \Omega$ is associated with a coefficient vector $\boldsymbol{\alpha}_i \in \mathbb{R}^k$. Specifically, the optimization problem is reformulated as follows:
\begin{equation} \label{equation:opt_sss_reformulated}
    \begin{aligned}
        & \underset{\boldsymbol{\alpha}_1, \ldots, \boldsymbol{\alpha}_N, S \subset \Theta}{\text{minimize}}
        & & \underset{i=1}{\sum^N }\ w(x_i) \left(\boldsymbol{\alpha}_i' C_S \boldsymbol{\alpha}_i - 2 \boldsymbol{b}_{x,S}' \boldsymbol{\alpha}_i + \phi(x_i,x_i)\right)\\
        & \text{subject to}
        & & |S| = k,
    \end{aligned}
\end{equation}
Notice that Equation \eqref{equation:opt_sss_reformulated} is a convex function in the coefficients $\boldsymbol{\alpha}_1, \ldots, \boldsymbol{\alpha}_N$. 

The final task is to set up decision variables for the measurement locations. Recall the definition of the measurement set $\Theta = \{y_1, \ldots, y_M\}$. Let $z_1, \ldots, z_M \in \{0, 1\}$ be binary decision variables with $z_i = 1$ if $y_i$ is selected, 0 otherwise. However, the objective function in its current form is not suitable for optimization as it is a function of both the set $S$ and coefficients $\boldsymbol{\alpha}_i$. The goal is to write it as quadratic function of the coefficient vectors \emph{only}. We can achieve this by replacing the set $S$ with the measurement set $\Theta$ in the objective and adding a constraint that allows a coefficient to be non-zero only if the corresponding measurement is selected. This is seen more clearly in the second last constraint in the following program for \textsc{\sss}:
\begin{equation} \label{equation:opt_sss_mip}
    \begin{aligned}
        & \underset{\substack{\boldsymbol{\alpha}_1, \ldots, \boldsymbol{\alpha}_N \in \mathbb{R}^M\\ z_1, \ldots, z_M}}{\text{minimize}}
        & & \underset{i=1}{\sum^N }\ w(x_i) g_{x_i}(\boldsymbol{\alpha}_i)\\
        & \text{subject to} & & \sum_{i=1}^M z_i = k,\\
        &&& (1-z_i) [\boldsymbol{\alpha}_j]_i = 0, \; \forall i, j,\\
        &&& z_i \in \{0, 1\}, \; \forall i,
    \end{aligned}
\end{equation}
where
\begin{equation} \label{equation:g_quadratic}
    g_x(\boldsymbol{\alpha}) := \boldsymbol{\alpha}' C_{\Theta} \boldsymbol{\alpha} - 2 \boldsymbol{b}_{x,\Theta}' \boldsymbol{\alpha} + \phi(x,x).
\end{equation}
 The second last constraint is a Special Ordered Set (Type 1) \cite{bertsimas2005optimization} and allows only one of $1-z_i$ and $[\boldsymbol{\alpha}_j]_i$ to be non-zero.

We now have our formulation. We transformed the problem in \eqref{equation:opt_sss_initial} to \eqref{equation:opt_sss_mip} by noticing the estimation error arises from the optimal linear estimator. The problem in \eqref{equation:opt_sss_mip} is now optimizing over the class of linear estimators as the optimization is over the coefficients $\boldsymbol{\alpha}_1, \ldots, \boldsymbol{\alpha}_N$.
The problem in \eqref{equation:opt_sss_mip} is a mixed integer quadratic program (MIQP). It has $M$ binary integer variables and $M \times N$ continuous variables. Though the number of decision variables from \eqref{equation:opt_sss_initial} has increased from $M$ to $M \times (1 + N)$, we now have a formulation that can be tackled by MIP solvers.

It is worth noting that the reformulation of the estimation error is not restricted to solving \textsc{\sss}; it is quite general. The estimation error can be written as a function of continuous and binary integer variables, which allows for formulations as MIQPs. In addition, there is no restriction on the covariance structure of the random field. This is a desirable benefit as physical processes exhibit complex variations on a multitude of spatial scales \cite{webster2007geostatistics}. Distinct regions of the environment can be modeled with different covariance structures. Further, we can also harness the anytime property of mixed integer solvers. We can terminate the solver at any point and obtain an approximate solution with a sub-optimality certificate.

\subsection{Informative Path Planning (\textsc{\ipp})} \label{subsection:informative_path_planning}
Recall the objective in informative path planning is to compute a budgeted path from a start to an end vertex in a directed graph $G=(\Theta,A,d)$. The vertices on the selected path must yield minimum estimation error. In this section, we will describe the MIP for informative path planning.

The setup requires additional decision variables to encode the path. We modify the setup in \cite{pataki2003teaching} used for the traveling salesman problem to account for orienteering constraints. Specifically, we define the binary integer variables $z_{ij} = 1$ if arc $(i,j)$ is on the path, 0 otherwise. Further, we ensure each the in-degree of each node (apart from the start and end node) is equal to its out-degree which can be at most 1. Similar to the MIQP defined in Section \ref{subsection:sparse_subset_selection}, we define $N$ continuous vectors $\boldsymbol{\alpha}_1, \ldots, \boldsymbol{\alpha}_N \in \mathbb{R}^M$ to model the coefficients. We assume that the start and end nodes are at indices $1$ and $M$. The following is a MIQP formulation.
    \begin{align} 
        & \underset{\substack{\boldsymbol{\alpha}_1, \ldots, \boldsymbol{\alpha}_N, \boldsymbol{z} }}{\text{minimize}} & & \underset{i=1}{\sum^N }\ w(x_i) g_{x_i}(\boldsymbol{\alpha}_i) \label{equation:opt_ipp_mip} \\
        & \text{subject to} & & \sum_{i=1}^M \sum_{j=1}^M z_{ij} d_{ij} \leq B, \label{constraint:path_length}\\
        &&& (1-\sum_{j=1}^M z_{ij}) [\boldsymbol{\alpha}_j]_i = 0, \; i=2,\ldots, M-1, \forall j \label{constraint:SOS}\\
        &&& \sum_{i=2}^M z_{1i} = \sum_{j=1}^{M-1} z_{jM} = 1 \label{constraint:start_end}\\
        &&& \sum_{j=2}^{M} z_{ij} = \sum_{k=1}^{M-1} z_{ki} \leq 1, \; i=2,\ldots,M-1, \label{constraint:in_degree}\\
        &&& \sum_{i,j \in S} z_{ij} \leq |S| - 1, \; \forall S \subset \Theta, |S| > 1, \label{constraint:subtour}\\
        &&& z_{ij} \in \{0, 1\}, \; \forall i,j
    \end{align}
where $g_{x_i}(\boldsymbol{\alpha}_i)$ is given by Equation \eqref{equation:g_quadratic}. 

We now describe the objective and constraints of the problem. The objective \eqref{equation:opt_ipp_mip} is to minimize the total estimation error. Constraint \eqref{constraint:path_length} requires the path length to be within budget. Constraints in \eqref{constraint:SOS} allow coefficients to be non-zero only if the corresponding measurement is selected. Constraints in \eqref{constraint:start_end} ensure the path starts at node $x_1$ and ends at node $x_M$. Constraints \eqref{constraint:in_degree} ensure connectivity of the path and that every node is visited at most once. The last set of constraints in \eqref{constraint:subtour} prevent subtours in the path. However, there are an exponential number of these constraints. As a result, not all the constraints are put into the solver at the beginning; the subtour elimination constraints (SEC) are implemented as lazy constraints. As the solver produces candidate integer solutions to \eqref{equation:opt_ipp_mip}, the violated subtour constraints are added to the program. The idea behind SECs is that most lazy constraints are unlikely to be violated so it is not necessary to generate them upfront. For more details, see \cite{pataki2003teaching}. 

\begin{rem}[Anytime Property \& SECs]
In Section \ref{subsection:sparse_subset_selection}, we discussed the benefit of the anytime property of the solvers. However, since we are implementing SECs as lazy constraints, it is possible that the solver produces an infeasible solution (subtours) if terminated early. This is unlikely to occur if the solver is given sufficient time. If the anytime property is a strict requirement, there are alternative approaches such as the MTZ formulation \cite{miller1960integer} that will always produce feasible solutions.
\oprocend
\end{rem}

\section{Evaluation} \label{section:results}

We provide empirical evidence of the advantages of our approach over branch and bound (\textsc{BnB}) \cite{binney2012branch} methods for informative path planning. Note that both approaches are global optimization methods. The first advantage is that when optimality is a requirement, the MIP computes solutions faster than \textsc{BnB}. Further, the MIP can compute solutions on instances where \textsc{BnB} is deemed intractable. Second, when both methods are given the same amount of computational resources, the MIP finds solutions of higher quality.

\begin{rem}[]
Even though we have a MIP for \textsc{\sss}, we focus on experimental evaluations of the MIP for \textsc{ipp}. This is because greedy algorithms for subset selection are efficient and known to perform well in practice \cite{hastie2020best,miller2002subset}.
\oprocend
\end{rem}

\emph{Experimental Setup}: We consider two environmental setups: grid-based graphs and probabilistic road maps. 
\subsubsection{Grid-Based Graphs}
We follow the setup in \cite{binney2012branch} where a Gaussian Process is indexed over a 2D environment i.e.,\ $D \subset \mathbb{R}^2$. The covariance structure of the field is modeled by the stationary squared exponential covariance function $\phi_{\text{SE}}: \mathbb{R}_{\geq 0} \rightarrow \mathbb{R}_{\geq 0}$ defined below:
\begin{equation}
     \phi_{\text{SE}}(h) := \sigma_0^2 e^{-\frac{h^2}{2L^2}},
\end{equation}
where $\sigma_0 = 1.0$ and $L=1.0$. The interpretation of $L$ is that it is roughly the distance one has to move in the field before the function value changes \cite{williams2006gaussian}. The graph is a $N \times N$ grid where each edge has length 1. The vertices of the graph form the observation set. The size of the prediction set is fixed at $M=25$. In all experiments, the start and end nodes are on diagonally opposite ends of the grid.

\subsubsection{Probabilistic Roadmaps}
We consider probabilistic road maps generated in a random field fitted to real world data. The dataset is collected from Broom's Barn, an 80 hectare farm in Suffolk, United Kingdom. It contains 435 potassium values sampled at 40 metre intervals on a 720 $\times$ 1240 grid. Our formulation only requires the covariance structure of the field to be specified which has been estimated in \cite[Section 8.7]{webster2007geostatistics}. The authors use the stationary spherical covariance function $\phi_{\text{sph}}: \mathbb{R}_{\geq 0} \rightarrow \mathbb{R}_{\geq 0}$ defined below:
\begin{equation}
\phi_{\text{sph}}(h) := 
    \begin{cases}
        c \left(1 - \frac{3h}{2a} + \frac{1}{2} \left(\frac{h}{a}\right)^3\right), & h \leq a\\
        0, & h > a,
    \end{cases}
\end{equation}
where the parameters are: $c=0.01519$ and $a=439.2$ metres. The graph is a probabilistic road map with a connection factor of 8 in the environment $D := \{ (x,y) \in \mathbb{R}^2: 0 \leq x \leq 720, 0 \leq y \leq 1240\}$, where the number of vertices is $M=100$. Figure~\ref{figure:intro_figure} depicts an example of a probabilistic roadmap (yellow edges).

In the following sections, The results are averaged over 5 runs where in each run, the prediction set and associated weights are generated uniformly at random. The experiments are implemented in NumPy \cite{harris2020array} and Gurobi \cite{gurobi} on an AMD Ryzen 7 2700 processor.

\subsection{Run Time}
\begin{figure}[t]
    \centering
    \includegraphics[width=0.99\linewidth]{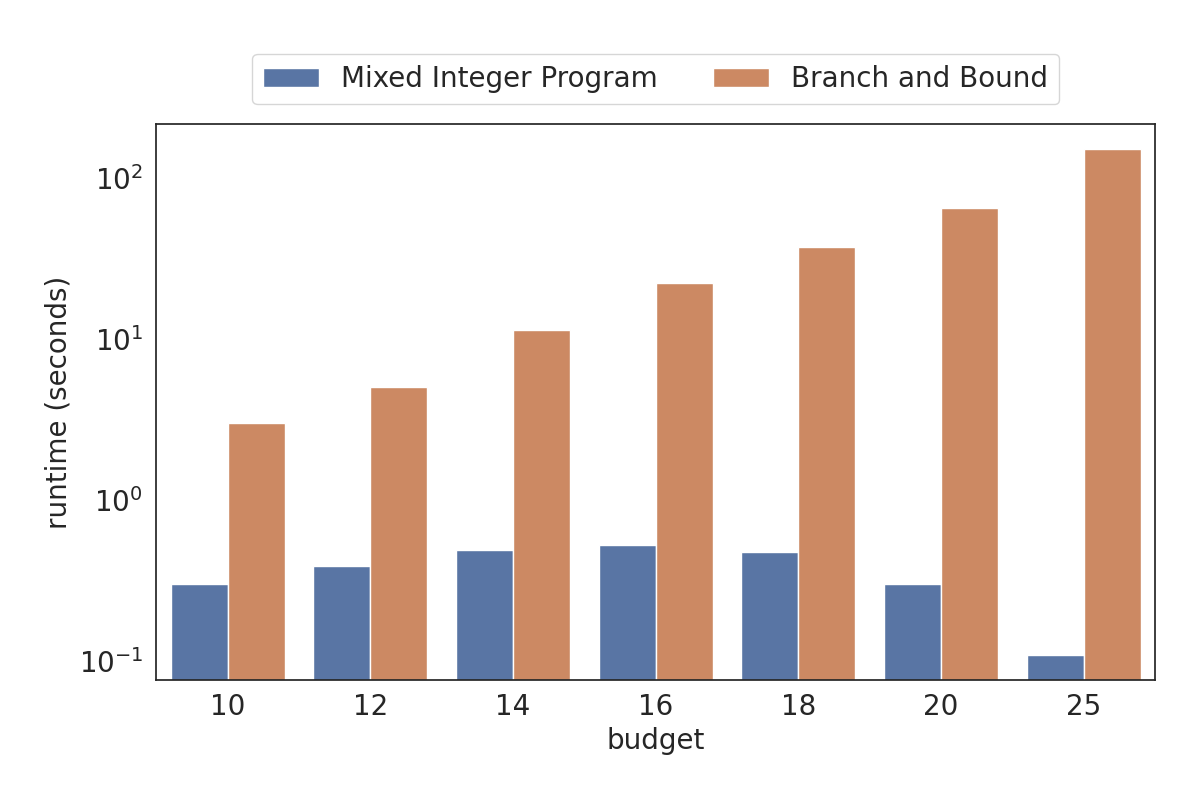}
    \caption{Running time (log scale) versus budget on path length of the MIP and \textsc{BnB} on a $5 \times 5$ graph. The MIP solves these instances in under a second while the runtime of \textsc{BnB} quickly becomes intractable.}
    \label{fig:ipp_exp1}
\end{figure}

We aim to answer the following question: when searching for optimally informative paths, which approach finds solutions faster? Since \textsc{BnB} can only compute optimal solutions to $5 \times 5$ grid-based graphs in a reasonable amount of time, we restrict the experiments for this subsection in this setting. We vary the budget on the path length from 10 (minimum feasible path length) to 25. The results are reported in Figure~\ref{fig:ipp_exp1}. The first observation is that the runtime of \textsc{BnB} grows rapidly as the budget length is increased. In contrast, notice that the MIP is able to solve all instances in under a second. An interesting observation is that the runtime increases until a budget of 16 and then decreases again. One reason for this might be that when the budget is high, there are few paths that visit all nodes (yielding minimum estimation error) within budget and the MIP is able to efficiently find them. These results indicate that \textsc{BnB} is limited to solving small instances and in that regime, the MIP can find solutions extremely quickly.

\begin{figure}[t]
    \centering
    \includegraphics[width=0.99\linewidth]{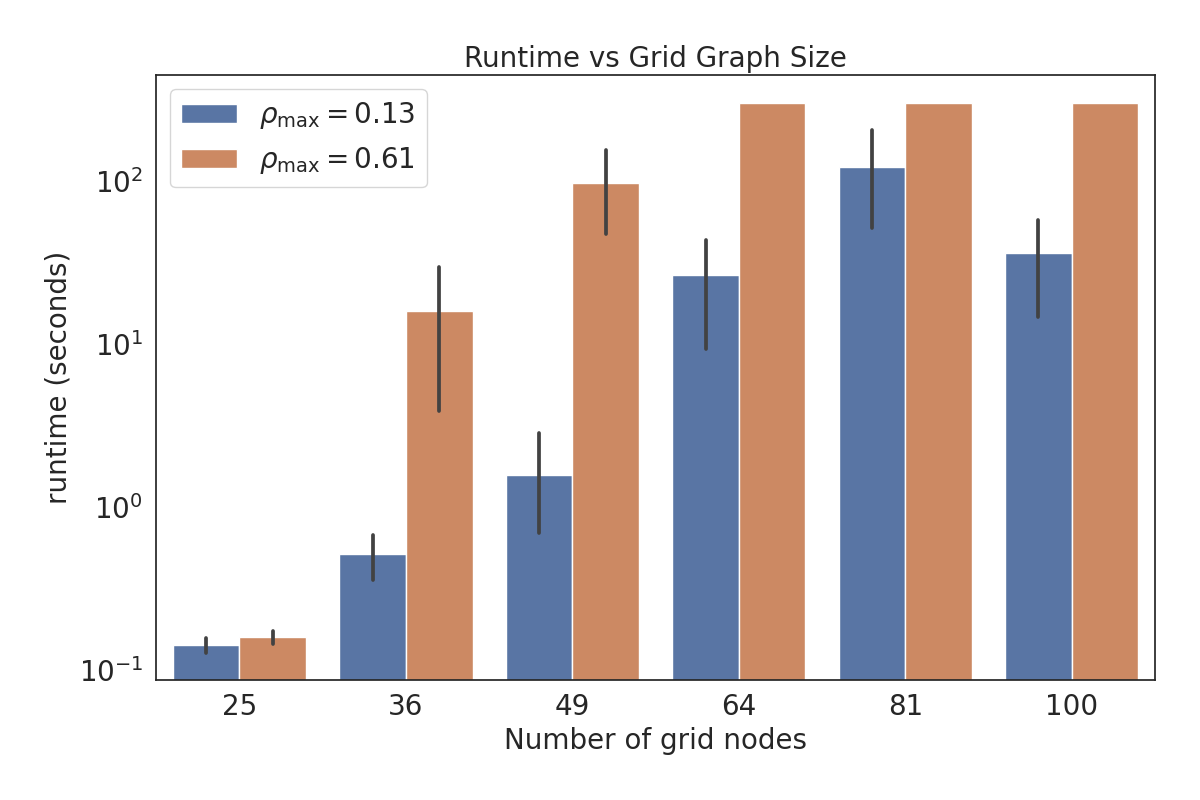}
    \caption{Runtime (log scale) of the MIP on larger instances of grid-based graphs. Highly correlated environments lead to long run times but in low correlation environments, the MIP performs well.}
    \label{fig:ipp_exp2}
\end{figure}

In Figure~\ref{fig:ipp_exp1}, the MIP solved tractable instances for \textsc{BnB} rather quickly. A natural question is: what is the set of tractable instances in grid-based graphs for the MIP? The answer depends on the maximum correlation $\rho_{\max}$ between any two distinct nodes in the graph as well as the number of nodes. We can control the maximum correlation between nodes through the parameter $L$ in the exponential covariance function. We consider two environments: $L=1$, corresponding to $\rho_{\max} = 0.61$ and the second with $L=0.5$, corresponding to $\rho_{\max} = 0.13$. For each environment, we increase the number of nodes in the grid-based graph from 25 to 100 and plot the run time for the MIP in Figure~\ref{fig:ipp_exp2}. Since the MIP can have long runtimes for certain graphs, we set a timeout to 2 minutes. In the case of high correlated environments, we see that the solver begins to timeout starting at $8 \times 8$ grids. However, when we consider the low correlation environments, the MIP seems to scale reasonably well even upto $10 \times 10$ grids. One reason for this might be that in high correlation environments, there are multiple good solutions since most grid points will yield low estimation error. Thus, it may be harder for the MIP to prune solutions. These results indicate that size of the graph alone does not determine the scalability of the MIP; the correlation of the field values has an important role to play.

\subsection{Solution Quality}
We switch our attention to solution quality. We aim to answer the following question: given the same computational budget in terms of run time, which approach finds paths that yield lower estimation error? We discuss results based on experiments in grid-based graphs and probabilistic roadmaps.

\begin{figure}[t]
    \centering
    \includegraphics[width=0.99\linewidth]{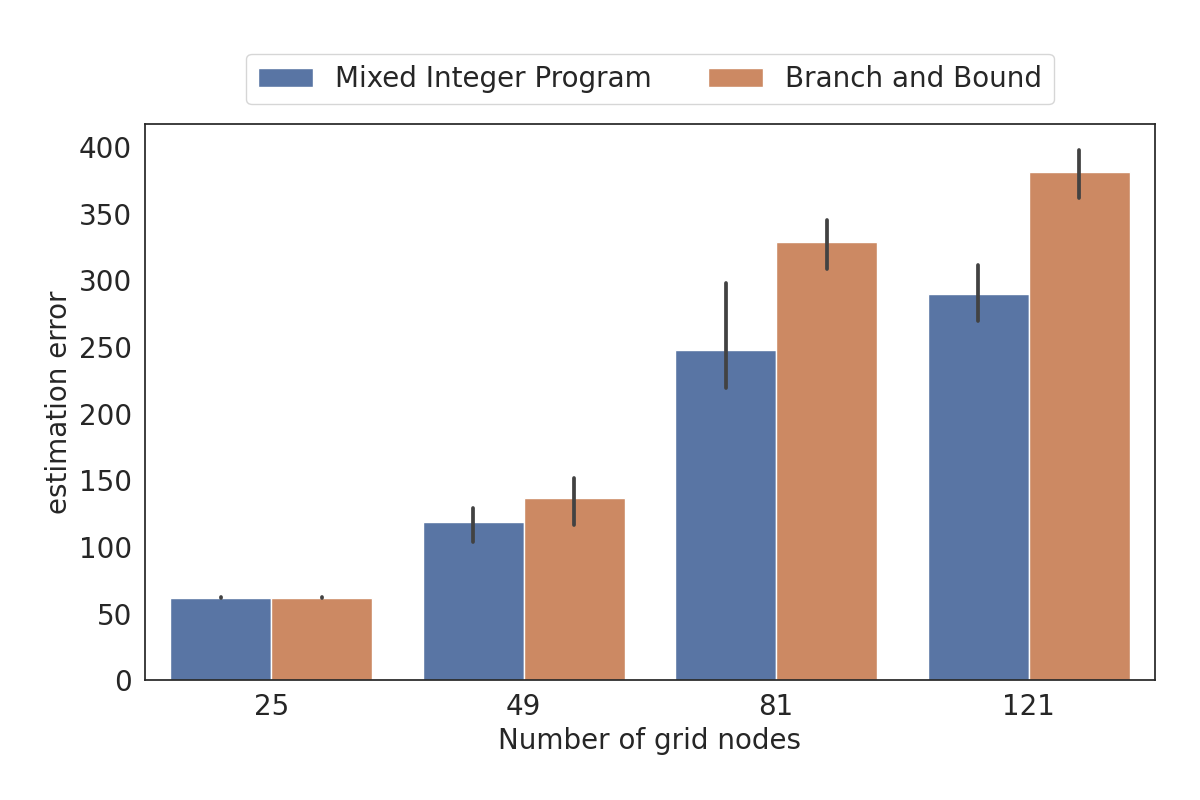}
    \caption{Comparison of the estimation error on large grid-based graphs. The x-axis denotes the number of nodes in the grid. The solvers are timed out after 2 minutes and the solutions are compared.}
    \label{fig:ipp_exp3}
\end{figure}

\begin{figure}[t]
    \centering
    \includegraphics[width=0.99\linewidth]{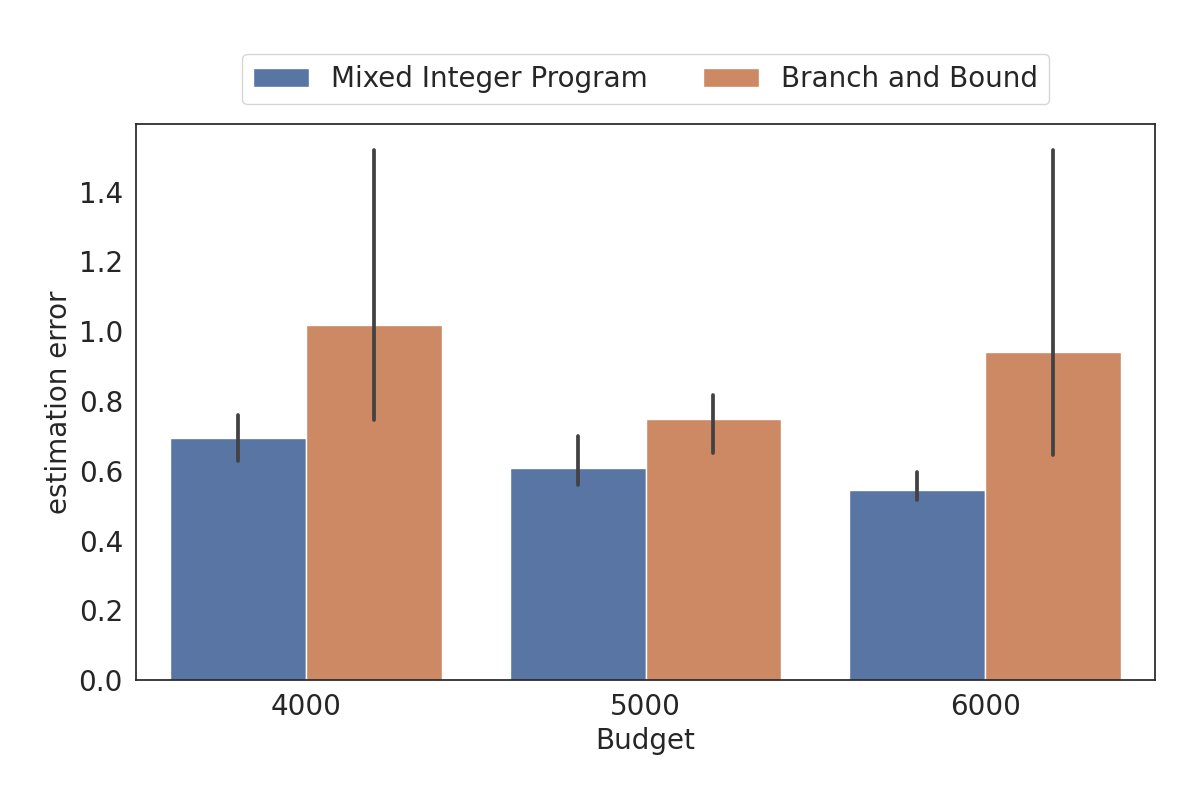}
    \caption{Comparison of the estimation error on probabilistic roadmaps. MIPs return lower estimation errors as the budget on the path length increases.}
    \label{fig:ipp_exp4}
\end{figure}

In grid-based graphs, we vary the number of nodes from 25 to 121 and report the estimation error returned from both \textsc{BnB} and the MIP. Since the running times can be quite high, we set a timeout to 2 minutes for each solver and return the best solution computed. The quality of those solutions, in terms of the estimation error, is reported in Figure~\ref{fig:ipp_exp3}. We see that the MIP attains lower estimation error across all graph sizes. In addition, as the size of the graph increases, the difference in error between MIP and \textsc{BnB} solution increases. These results indicate that even though the MIP has a long running time, it returns approximate solutions of better quality when compared to \textsc{BnB}. A similar result holds in the experiments with probabilistic roadmaps. Since the roadmaps contain more edges than grid-based graphs, the task is more challenging and we expect much longer running times to find optimal solutions. Instead of varying the size of the graph, which is fixed to 100 nodes, we vary the budget on the path length instead. The time out for both solvers was set to 5 minutes and the results on the solution quality are reported in Figure~\ref{fig:ipp_exp4}. The estimation error obtained by the MIP is lower than the \textsc{BnB} across all budgets. As we increase the budget on the path length, the MIP obtains solutions of better quality. This is expected as the robot is able to take additional measurements. However, the branch and bound algorithm does not display the same behaviour. The error obtained increases and then decreases again. One reason for this might be because of the order of enumeration of paths by the \textsc{BnB} solver. In some cases, it may be able to find good approximate solutions but not always. The results of this indicate that in the of probabilistic roadmaps, the MIP solver returns better quality paths than \textsc{BnB}.

\section{Conclusions} \label{section:conclusions}

We discussed the problem of informative path planning in random fields with finite observation and prediction sets. We provided a new mixed integer program formulation of the problem. The key idea was to expand the search space to optimize over all linear estimators and the measurement subset. This enabled a mixed integer formulation which was convex in the continuous variables which allowed us to leverage the power of modern optimization solvers. Our results indicate this approach is appealing because it has shorter running times and better solution quality than previous branch and bound approaches. Looking forward, this could be extended in two ways. First, the structure of mixed integer program may be studied to design efficient approximation algorithms. Second, the formulations could be extended to handle multiple robots, tackling the multi-robot informative path planning problem.

\Urlmuskip=0mu plus 1mu
\bibliographystyle{IEEEtran}
\bibliography{mybib}
\end{document}